\documentclass[11pt,twoside]{article}
\usepackage{asp2004}
\usepackage{psfig}
\usepackage{epsf}
\usepackage{graphics}
\usepackage{lscape}
\begin{document}
\title{Numerical Simulations of the MRI and Real Disks}
\author{Steven A. Balbus}
\affil{\'Ecole Normale Sup\'erieure, Laboratoire de
Radioastronomie, 24, rue Lhomond, 75231 Paris, France \& \\
Astronomy Dept., Univ. Virginia, Charlottesville, VA 22903, USA}

\begin{abstract}
Numerical MHD codes have become extraordinarily powerful tools with
which to study accretion turbulence.  They have been used primarily to
extract values for the classical $\alpha$ parameter, and to follow
complex evolutionary development.  Energy transport, which is at the
heart of classical disk theory, has yet to be explored in any detail.
Further topics that should be explored by simulation include nonideal
MHD, radiation physics, and outburst behavior related to the temperature
sensitivity of the resistivity.
\end{abstract}

\section{Introduction: The Need for MHD Simulations}

Magnetic fields, particularly weak magnetic fields, strongly destabilize
differential rotational in astrophysical accretion disks, allowing them
to live up to their name.  The magnetorotational instability (MRI) is but
one manifestation of a very general behavior of weak magnetic fields in
plasmas: by forcing angular momentum and entropy transport to follow field
lines, they allow free energy gradients (temperature and angular velocity)
to become sources of instability.  Despite the fact that they have only
been recently understood in detail, the instabilities are not difficult
to simulate and to study numerically.  (A single processor workstation
will do.)  For this reason, numerical investigations of spontaneously
arising, undriven accretion disk turbulence are based on the MRI.

Nevertheless, the notion that simple differential rotation is
intrinsically unstable via nonlinear {\em hydrodynamic} instability
(Dubrulle 1993) continues to attract adherents (e.g., Richard \& Zahn
1999; Richard 2003), in part because there is as yet no mathematical
proof of nonlinear stability.  It is argued that shear layers are known
to be linearly stable, but extremely sensitive to nonlinear disruption,
thus Keplerian differential should also be nonlinearly unstable, provided
that the Reynolds number is sufficiently high.

In fact, one does not even have to resort to nonlinearity to find
an example of shear induced instability.  {\em Linear} global hydro
instabilities can be present under some circumstances, provided that
proper boundary conditions are specified.  These ``Papaloizou-Pringle"
(Papaloizou \& Pringle 1984; Goldreich, Goodman, \& Narayan 1986)
instabilities, which involve communication by trapped waves on either
side of a corotation radius, can disrupt systems that would ordinarily
be stable by the classical Rayleigh criterion of increasing outward
specific angular momentum.  Numerical study of the PP instability poses no
unresolvable difficulties, and it may be followed well into its nonlinear
regime (Hawley 1991).  Indeed, the emergent characteristic behavior of
this instability is an instructive example of how a controlled simulation
can be used in pursuing an answer to a well-posed problem.  The PP
instability creates spiral structure in the disk, rather than turbulence,
since this turn out to be a more efficient angular momentum transport
mechanism in a hydrodynamical flow.  But both because the instability
is very sensitive to the proper boundary conditions, and because the
instability weakens dramatically when the pressure gradients vanish,
the relevance PP instabilities to Keplerian disks is far from clear.

More to the point, numerical simulations offer no support for the
notion that local differential rotation is itself nonlinearly unstable.
Proponents of nonlinear instability argue that this is simply because
existing codes do not have a high enough effective Reynolds number,
and are thus too diffusive.  I must confess to being somewhat baffled
by this argument, since the nonlinear instability of a shear layer is
readily demonstrated, and it certainly does not require outlandishly
high Reynolds numbers.  Moreover, tight convergence has been achieved
between codes with very different diffusive properties, and these
show unambiguous nonlinear stability of local Keplerian flow (Hawley,
Balbus, \& Winters 1999).


It has been recently noted that the local ``shearing box''
equations of motion (e.g. Balbus \& Hawley 1998), which are fully
nonlinear but do not include radial curvature terms, exhibit an
exact scale invariance (Balbus 2004b).  Specifically, if ${\bf w}
({\bf r}, t)$, $P({\bf r}, t)$, and $\rho ({\bf r}, t)$ represent
exact solutions for the velocity, pressure and density
respectively, then
$$
{1\over\beta} {\bf w} (\beta {\bf r}, t), \quad \rho(\beta {\bf r}, t),
\quad
{1\over\beta^2} P(\beta {\bf r}, t)
$$
where $\beta$ is an arbitrary scaling factor, are also exact solutions.
This means that Keplerian flow is no more unstable at very small scales
then at larger, computationally accessible scales.  Any putative local
disk instability in Keplerian flow would have to exhibit behavior very
different from that seen in the classical breakdown of planar shear flow.
Clearly, Coriolis forces are strongly stabilizing, rigorously so in the
linear regime, and it is not surprising that the stabilization extends
into the nonlinear regime as well.  As we have already noted, this does
not mean that any rotationally-based instability is proscribed, it simply
means that {\it local} shear is insufficient to disrupt Keplerian flow.
Nonlinear instability buffs are forced to argue that conditions beyond
the reach of modern supercomputers are required to trigger instability,
a position that may become increasingly untenable as time goes on.
As a very practical matter, the MRI remains the only instability on the
market that manages to sustain long term, unforced turbulence in numerical
simulations of Keplerian disks.  In the end, numericists really don't
have a choice.

\section {Tools of the Trade: Correlated Fluctuations}

Accretion disk turbulence cannot yet be directly observed in nature,
and in the laboratory the MRI has only very recently been identified
(Sisan et al. 2004).  Three-dimensional local MHD simulations are
almost ten years old at the time of this writing (Hawley, Gammie, \&
Balbus 1995).  Three-dimensional global simulations followed shortly
thereafter (Armitage 1998; Hawley 2000, 2001).  In fact, two-dimensional global
simulations date back two decades (Uchida \& Shibata 1985), well before
MRI calculations appeared in local guise, and before the MRI had been
identified in accretion disks.   By now, global MHD simulations are a
small industry, with several groups around the world investigating many
different specialized problems.  What can be learned from these efforts?

We must first understand what it is that we don't understand, and for
that we should begin with the tenets of classical disk theory (CDT).
CDT is more widely known as viscous disk theory, a moniker that surely
needs to be laid to rest.  CDT helped to orient our intuitions during
the early stages of disk theory, but now that this phase has past, it is
time for disk viscosity to be jettisoned.  The analogy has became more
of a hindrance than anything else --- in some papers enhanced viscosity
and turbulence are viewed as competing processes!  It is time for disk
theorists to move out of the 1970s, and relegate ``anomalous viscosity''
to the cultural dustbin along with afros, leisure suits, and disco.

\subsection {Stress Tensor}

The fundamental transport quantity that is at the center of any
theory of turbulent disk transport is the stress tensor.  Unlike a
viscous stress, the true stress tensor is entirely a property of
the flow, not of the constituent fluid.  It is defined in terms of
correlated velocity fluctuations and correlated magnetic fields.
Let ${\bf u}$ be the {\em difference} between the local disk
velocity and pure Keplerian rotation on cylinders.  Let ${\bf
u_A}$ be the Alfv\'en velocity associated with the magnetic field
${\bf B}$, and mass density $\rho$,
\begin{equation}
{\bf u_A} = { {\bf B}\over \sqrt{4\pi\rho} },
\end{equation}
and is not necessarily a fluctuating quantity. We assume that the
magnitudes of ${\bf u_A}$ and ${\bf u}$ are comparable, and that
both are much less that the unperturbed rotation velocity,
$R\Omega$, where $R$ is the cylindrical radius. The density and
pressure fluctuations, $\delta\rho$ and $\delta P$ are assumed to
be small compared with their mean values, $\bar \rho$ and $\bar
P$, respectively. The stress tensor has, in principle, six
independent components, but in practice only the azimuthal
components play an essential role in disk theory,
\begin{equation}
{\bf T} = \bar\rho \langle ({\bf u} u_\phi - {\bf u_{A}}
u_{A\phi}) \rangle
\end{equation}
and we shall henceforth refer to these components as {\em the}
stress tensor. Extracting an accurate mean value of ${\bf T}$ from
numerical simulations requires patience, as very long time
averaging may be required (Steinacker \& Papaloizou 2002).

The stress tensor plays two mathematically related, but physically
distinct roles is disk theory: (1) it is directly proportional to
the outward angular momentum flux through the fluid, and hence to
the classical $\alpha$ parameter; and (2) it is directly
proportional to the rate at which energy is extracted from the
background differential rotation of the disk.  The latter is the
free energy source that powers the turbulent fluctuations
themselves.  The central tenet of CDT is that this energy goes
nowhere.  Instead, it is all locally dissipated and radiated away
on the spot.  The reader will recall the well-known factor of 3
discrepancy between the radiated energy and the local orbital
energy that is released.  This is usually explained as a nonlocal
consequence of the presence of an energy flux, but I prefer to
think of it as the result of the local injection of the free
energy of differential rotation.  In any case, this assumption of
local energy dissipation is very powerful, and needs to be
carefully checked.

\subsection {Fluctuation Equations}

\subsubsection{Hydrostatic equilibrium}
The steady state disk equations have been worked out under
assumptions of section 2.1.
The fundamental equation of hydrostatic equilibrium is
\begin{equation}\label{hse}
R\Omega ^2 {\bf e_R} = {1\over {\bar \rho}}\nabla {\bar P} + \nabla\Phi
\end{equation}
where ${\bf e_R}$ is a unit vector in the radial direction,
and $\Phi$ is the central potential.  Notice that hydrostatic equilibrium may be
expressed in terms of mean values, and is not directly affected by the
turbulent fluctuations.  Equation (\ref{hse}), and those to follow,
are two-dimensional: the process of extracting the mean averages out
the azimuthal structure.  Hydrostatic equilibrium breaks down, of course,
in regions of thermal or transonic flow.

\subsubsection{Mass conservation}
By way of contrast, transport processes are intimately linked to correlations
in fluctuating quantities, and to the magnetic field.  (This is true even if
the field is weak.)  The mass flux is
\begin{equation}\label{mflux}
\langle \rho {\bf v} \rangle = {\bar\rho}\bar{\bf v} +
\langle \delta \rho \delta {\bf v} \rangle,
\end{equation}
and in steady state
\begin{equation}
\nabla \cdot \langle \rho {\bf v} \rangle =0
\end{equation}
The final fluctuating term in equation (\ref{mflux}) is generally
comparable to the mean flow term, though it is dropped
in viscosity analogues.

\subsubsection{Angular momentum}

Angular momentum conservation is expressed by the steady state
equation
\begin{equation}
\nabla\cdot \langle R^2 \Omega
\langle \rho {\bf v}\rangle +  R {\bf T}\rangle = 0
\end{equation}
The first term represents that angular momentum transported directly
by the mass flux, whereas the second is flux passing through the disk
gas.  In CDT, only radial transport is considered, and
${\bf T}$ is replaced by a viscous stress.

\subsubsection{Energy conservation}

Energy is transported through the body of the disk
via the thermal energy flux,
\begin{equation}
{\bf F_{th}} = {\gamma\bar\rho \over \gamma - 1}\langle \delta {\bf v}
\delta \tau \rangle
\end{equation}
where $\gamma$ is the adiabatic index, and $\tau$ is the normalized
temperature $kT/\mu$ ($k$ is the Boltzmann constant and $\mu$ the mean mass
per particle).  This form of $F_{th}$ is very general.  The velocity-temperature
correlation is responsible for convective transport, as well as the energy
transported by ordinary sound waves.
The energy equation for a turbulent disk is (Balbus 2004a):
\begin{equation}\label{energy}
- T_{R\phi} {d\Omega\over d\ln R}
=
\nabla \cdot {\bf F_{th}} + {\bar\tau}\langle\rho{\bf v} \rangle \cdot
\nabla S +Q_{rad}
\end{equation}
where $T_{R\phi}$ is the radial component of ${\bf T}$,
$Q_{rad}$ is the volumetric radiative loss rate, and $S$ is in essence
the mean entropy,
$$
S\equiv {1\over\gamma-1} \ln {\bar P \over {\bar\rho}^\gamma}
$$
Equation (\ref{energy}) states that the rate at which
energy is locally extracted from the differential rotation is equal
to the rate at which it is carried away by waves, dissipated as heat,
and radiated away.

In CDT, all terms on the right side of the equation
are ignored except for radiative losses, $Q_{rad}$.
This is justified in the following sense.  It is customary
to express the stress tensor in terms of a simple dimensionless
scaling,
\begin{equation}
T_{R\phi} = \alpha P
\end{equation}
where $\alpha$ dimensionless parameter that
embodies both the magnitudes of the velocity fluctuations as well as
their relative degree of correlation.  This will of course be immediately
recognized as the relationship that gives ``$\alpha$-disk'' theory its name.
Since there is more than one correlation tensor of interest here, let
us relabel the above alpha as $\alpha_L$, since $T_{R\phi}$ is associated with
angular momentum transport.  We are also free to invoke a scaling law of the
form
\begin{equation}
F_E = \alpha_E P{\bar\tau},
\end{equation}
where $F_E$ is a radial energy flux. The neglect of this thermal
energy flux will then be justified if
\begin{equation}
{\bar\tau\over R\Omega}{\alpha_E\over\alpha_L} \ll 1
\end{equation}
The first ratio is an inverse Mach number for the rotation
velocity, and will be small (by definition) for a thin Keplerian
disk.  This presumably includes CV disks.  The energy flux term
will then be negligible if $\alpha_E$ does not much exceed
$\alpha_L$.  While this is not a priori unreasonable, neither is
it very well tested.  It is the kind of question that should be
asked of numerical simulations.

\subsection {Global Simulations}

Most of the the global disk simulations to date follow the
following prototype.  The gas is a polytropic torus, in the
gravitational field of a ``black hole.''  This means that the
potential field is given by (Paczy\'nski \& Wiita 1980):
\begin{equation}
\Phi = -{GM\over |{\bf r - r_g}|},
\end{equation}
where $r_g$ is the usual Schwarzschild (or gravitational) radius,
$2GM/c^2$. (Symbols have their usual meanings.) This initial state
is chosen to be a torus because it allows a simple hydrostatic
equilibrium to be constructed, but it is also physically
reasonable: a reservoir of angular momentum bearing material
located at a large distance from the origin.  The equilibrium is
of course nonmagnetic; as soon as a small field is added, the MRI
quickly dominates the evolution.

The original motivation for introducing the Paczy\'nski-Wiita
potential was that it captured some key features of true black
hole dynamics in a user-friendly Newtonian format.  For the
numericist, this is a particularly convenient potential function,
since the inner boundary is pure outflow, or, from the point of
view of the hole, an inflow at $r=r_g$.  The use of this potential
when the central object is {\em not} a black hole is obviously a
cheat, but it does finesse the problem of how to handle the
numerically intractable disk-star boundary layer. The price to be
paid is that all information of this observationally critical
region is lost.  But if the scope of the simulation is restricted
to the study of the fundamental turbulent transport properties in
the body of the disk, this approach makes some sense, particularly
at this early stage. One must be willing to accommodate the
possibility, however, that the existence of the inner boundary
layer could cause global changes in ways that are difficult to
foresee.

If we take at face value the notion that numerical codes calculate
the evolution of a strictly polytropic disk, equation (\ref{energy})
leaves only one choice for its dominant balance:
\begin{equation}
- T_{R\phi} {d\Omega\over d\ln R}
=
\nabla \cdot {\bf F_{th}}
\end{equation}
This is a balance between the thermal energy flux divergence and
the rate at which energy is exchanged with differential rotation, a
purely adiabatic process.   In the case of WKB waves, it leads to the
conservation of wave action, rather than wave energy.  More generally,
this balance would involve extraction of the free energy of differential
rotation followed by its active transport through the disk.  This would
involve an adiabatic coupling between evanescent and wavelike modes.
Does this really happen in simulations, let alone real disks?

It hasn't been checked.  ZEUS-like codes, which base their
energetics on an internal, as opposed to a total energy equation,
tend to be lossy when computing small scale flow structure. This
is not necessarily bad, and in fact it mimics the behavior of real
disks.  In numerical tests of the stability of hydrodynamical
flow, these energy losses were tracked and their rapid growth used
as a hallmark for the presence of turbulence (Balbus, Hawley, \&
Stone 1996).  These losses introduce an effective $Q_{rad}$ term
in the calculation, even though it is not explicitly included in
the code.  The interesting question is how much free energy is
locally ``radiated'' and how much is transported elsewhere, since
this will determine whether local disk models make sense.  The
fact that simulations do not lead to very thin disks suggests that
local grid losses are not overwhelming, but a detailed accounting
should be part of the numerical diagnostics.  The existence of
thin disks as well as the local $R^{-3/4}$ scaling law seen in
some eclipsing binaries like {Z Cha} (Frank, King, \& Raine
2002) are certainly consistent with local dissipation, but it is
not obvious that they require it.

\section {Disk Morphology}

Disk morphology predictions may be more robust than is often
recognized. The angular velocity $\Omega$ tends to be constant on
cylinders, since $z$-dependent rotation profiles are unstable
(Goldreich \& Schubert 1967). If so, the right hand side of
equation (\ref{hse}) is a pure gradient, and whatever the actual
constitutive relationship between $P$ and $\rho$ may be, the
equilibrium functional relationship between these two quantities
will be that isobaric and isochoric surfaces coincide.  Moreover,
these surfaces are also equipotentials, provided that the
effective centrifugal potential is included (e.g., Frank et al.
2002).

In principle, the rotation profile $\Omega(R)$
is a free functional parameter, but in practice
almost all MRI simulations show that a Keplerian power law rapidly
emerges:
\begin{equation}
\Omega^2 = {GM\cos\beta\over R^3}
\end{equation}
where $\cos\beta$ is a proportionality constant not very different
from unity. The reason for this seems to be connected with the
vigorous outward angular momentum transport that always
accompanies the MRI in disks, which spreads the initial torus
radially, diluting the dynamical effects of pressure gradients in
the process.  The disk then seems to follow a simple a
quasi-Keplerian profile, linear in $GM$, with residual pressure
support allowing rotation at slightly below its pressure-free
value. For a polytropic equation of state, $P=K\rho^{\gamma}$, the
equilibrium density profile satisfies
\begin{equation}\label{adi}
{K\gamma\rho^{\gamma-1}\over\gamma-1} =
{\cal H_\infty} + {GM\over r}\left( 1 - {\cos\beta \over
\cos\lambda}\right)
\end{equation}
where ${\cal H_\infty}$ is the enthalpy at infinity, and $\lambda$
is the colatitude, $\pi/2 - \theta$, where $\theta$ is the usual
spherical angle.  An isothermal equation of state yields:
\begin{equation}\label{iso}
\rho = \rho_{\infty}\exp\left[ {GM\over r}\left( 1 - {\cos\beta \over
\cos\lambda}\right)\right]
\end{equation}

\begin{figure}[t]
\plotone{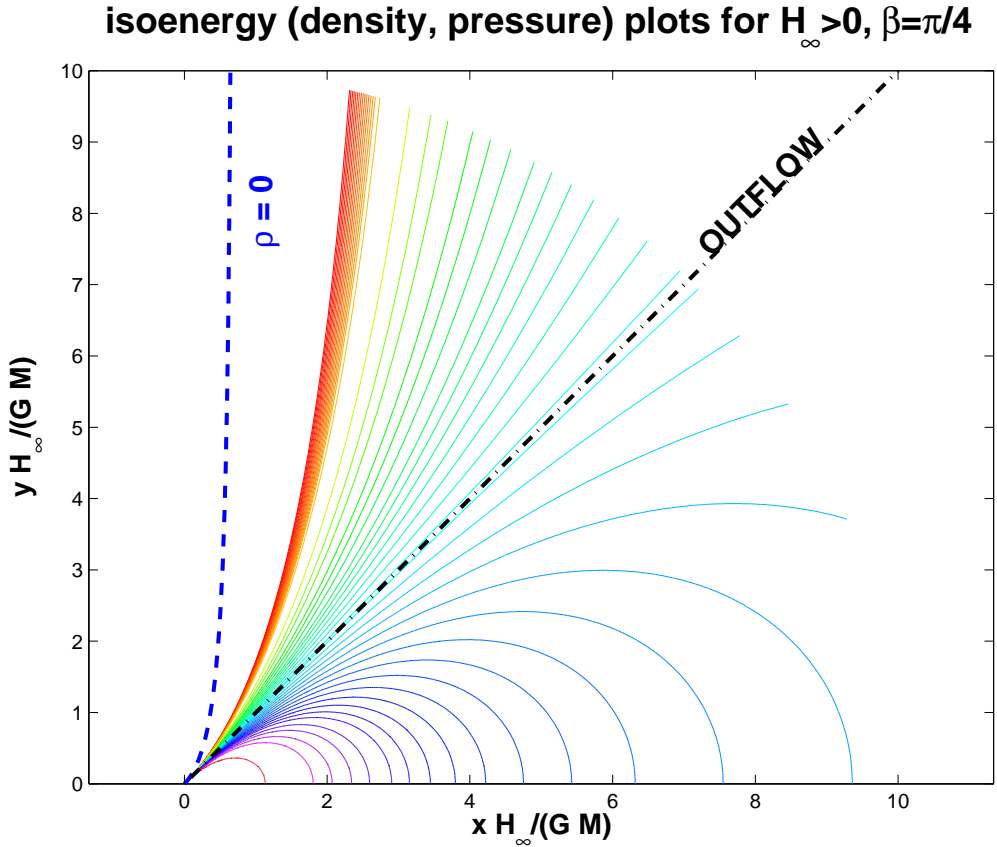}
\caption[]{Equipotential adiabatic disk contours for the case when $\Omega$
is 0.84 of its Keplerian value, corresponding to an opening wedge angle
of $45^\circ$, marked by the separatrix line marked OUTFLOW.
Open contours become very closely packed near
the $\rho=0$ boundary.}
\end{figure}

There is a qualitative difference between an adiabatic and an
isothermal disk.  An adiabatic disk has a well-defined
sharp edge at the value of $\lambda$
corresponding to $\rho=0$, whereas an
isothermal disk trails off exponentially above the disk boundary
$\lambda=\beta$.
The equipotential surfaces themselves
satisfy the equation
\begin{equation}
{1\over r}\left( 1 - {\cos\beta\over \cos\lambda} \right) = C= {\rm constant.}
\end{equation}

Contours for a fiducial example are shown in Figure 1.
When $C>0$, the contours are open, extending to infinity.  When $C<0$,
they curl back through the midplane and reconnect at the origin.
The critical $C=0$ contour is the straight line $\lambda=\beta$,
dividing the two classes.  Notice that all contours converge to this
critical contour as $r\rightarrow 0$.

The zone of ``contour convergence'' seems to be the site of the launching
of a jet-like outflow in numerical simulations, giving some credence
to this analytic model.  (The jet appears to be thermal, rather than
magneto-centrifugal, in origin.)  The tight grouping of the equipotential
surfaces in the convergence zone allows dissipative processes to move
fluid elements from one equipotential surface to another fairly easily
in this region.  Incoming disk material cannot be swallowed as fast as
it arrives, and is forced onto upward onto open equipotential surfaces.
The inner most allowable surface will always coincide more closely
to a constant angular momentum cylinder then will surface closer to
the disk.  This will be the easiest contour for the fluid to flow along
as it leaves the disk, requiring a minimum of angular momentum change.
And indeed, the simulations show a pile-up of material close to the last
open equipotential surface, confined against this surface from the inside
by a low density magnetic corona.  Significantly, the isothermal runs do
not show the formation of a jet.  This can be understood both because the
softer equation of state allows material to be more easily swallowed at
the center of flow, and because the density is exponentially curtailed
above the cone $\lambda=\beta$.

\begin{figure}[t]
\plotone{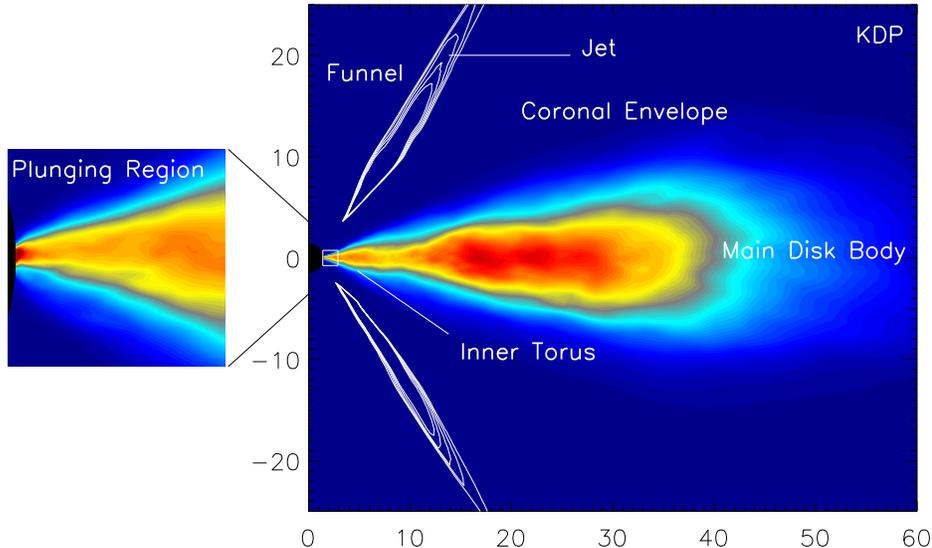}
\caption[]{Cross section of density contours of
an adiabatic accretion disk surrounding
a Kerr black hole. Contour shape and presence of jet-like
outflow follow equipotential surfaces from equation (17).}
\end{figure}

Figure 2 shows a detailed simulation by De Villiers, Hawley, \&
Krolik (2003).  The plot displays azimuthally averaged density
contours after a period of time corresponding to 10 orbital
periods at the pressure maximum of the initial torus used in the
simulation.  This is a general relativistic calculation of an
adiabatic gas, but except very near the ``plunging region'', the
gravitational potential is nearly Newtonian. The tear drop cross
section predicted by analytic theory and inner jets are both
present.

\section {Beyond Ideal MHD}

\subsection {Hall and Ohmic Processes}
At temperatures below 2000 K, the level of ionization in a CV disks can
fall below the minimum needed to keep the gas in the ideal MHD regime.
Assuming that dust grains are unimportant, the relative mutual drift
velocities of ions, electron, and neutrals gives rise to ambipolar
diffusion, Hall electromotive forces (HEMFs), and Ohmic dissipation.
Ambipolar diffusion is important in interstellar clouds, but for the much
denser disks of interest here, HEMFs and Ohmic losses are of primary
interest.  Numerical MHD has recently begun to study these nonideal
effects, which can have a decisive influence over disk behavior.

Ohmic losses were studied in local disk calculations
by Fleming \& Stone (2003), who found that MHD
turbulence could be maintained only if the magnetic Reynolds number
$$
Re_M \equiv {c_S^2\over \eta \Omega},
$$
where $\eta$ is the resistivity, exceeded a critical value, $Re_M(crit)$.
The value of $Re_M(crit)$ depends upon the field geometry.  For runs with a
mean vertical field present, Fleming \& Stone found critical values
of about 100; these increased to $10^4$ for runs without a mean vertical
field.  Once $Re_M < Re_M(crit)$, turbulence is quickly suppressed.
The addition of HEMFs (Sano \& Stone 2002) seems not to alter very much the
values of $Re_M(crit)$, but the saturations levels are more sensitive.
These can be significantly raised by the presence of the Hall effect.

Numerical non-ideal MHD offers yet another venue of opportunity.  The ohmic
resistivity is extraordinarily temperature sensitive in the transition
regime, because it depends upon the electron ionization fraction.
If thermal in origin, this fraction depends upon Boltmann factors of
the form $\exp(-I/kT)$ where the ionization energy $I$ is well above
$kT$, which accounts for the temperature sensitivity.  Small temperature
fluctuations cause large changes in the ionization fraction.  An increase
in $T$ leads to a decrease in $\eta$, and, it may be expected, to higher
levels of turbulence.  This is a prescription for yet higher temperatures,
and a possible runaway.  A similar instability is associated with $T$
decreases and suppression of turbulence.  How this resolves itself
is not yet known, but it seems to me that the problem is within the
computational realm.

Applications of this behavior to DN eruptions have been explored (Gammie
\& Menou 1998; Menou 2000; Sano \& Stone 2003), but interpreting the
quiescent stage in this context remains problematic.  Fleming \& Stone
(2003) make the interesting and provocative suggestion that magnetically
``dead'' zones may be affected by surrounding lower density magnetically
active regions.  In this view, the dead zones may be only morbid, and
host an active stress tensor maintained by outwardly transporting
trailing density waves.  If this is in fact the way that such low
ionization regions behave, it is a bit of a coup for numerical studies of
accretion disks.  It is difficult to imagine a compelling {\em a priori}
analytic argument, for example, that would give life to the dead zone
in this way.

The difficulty with the low state has been to avoid having the $\alpha$
parameter drop all the way to zero.  Gammie \& Menou (1998) suggested global
hydro-instabilities, Menou (2000) tidal interactions, and Sano \& Stone
(2003), in
their numerical calculation, treated the low state ``by hand''.  The
view put forth above offers another possibility: the quiescent disk
state might be a hybrid wave-turbulent system, while the high state is a flow
fully governed by the MRI.   In its low state, the disk is obstructed by
a sluggish but extended central zone, while accretion occurs more easily
in a rarified atmosphere around the low ionization region.  Even in the
low hybrid state, the primary accretion process may still be largely MHD,
but affecting only the tenuous gas.

\subsection{Radiation}

Radiation is essential to CDT, since this is the fate of all the free
energy that is locally dissipated by turbulence.   Very little work has
been done, however, to include radiative losses in numerical simulations.
What has been done, has focused on the case where the radiation energy
rivals or dominates the thermal energy density, which can happen
in the inner regions of black hole accretion.  The linear stability
of a magnetized, stratified, radiative gas has been investigated by
Blaes \& Socrates (2001).  Despite the complexity of the full problem,
the MRI emerges at the end of the day unscathed, its classical stability
criterion $d\Omega^2/dR > 0$ remaining intact.

The proper interpretation of a linear analysis is slightly unclear,
since the unperturbed state is already presumably fully turbulent due
to the MRI; a rotationally stable disk lacks an internal energy source.
The interplay between MHD turbulence and radiation is intrinsically
nonlinear.  Turner, Stone, \& Sano (2002) have studied a local,
radiative, axisymmetric, shearing box.  The linear calculations of Blaes
\& Socrates were confirmed in detail, and the nonlinear flow fully
developed.  As in standard MRI simulations, the stress is dominated
by the Maxwell component, which is a factor of a few larger than the
Reynolds terms.

Clearly, much could be learned by returning to the case where radiation
is an important loss mechanism, but does not dominate the local energy
density---i.e., the Shakura \& Sunyaev (1973) prototype.   The technical
expertise brought to bear on radiation dominated disks would serve this
simpler problem well.  Establishing that a radiative shearing box does,
in fact, evolve toward something resembling the classical Shakura-Sunyaev
paradigm would be an important milestone for accretion theory.

\section {Summary}

\begin{itemize}

\item The MRI remains the leading candidate for the origin of
enhanced angular momentum and energy transport in accretion disks.
A compelling case for sustained hydrodynamical turbulence has yet to be
made.

\item Despite very real limitations, the most serious of which is probably
still the limited dynamical range, numerical simulations have become
truly powerful.  They have allowed us to follow the global evolution
of polytropic disks, and to witness the build-up of a Keplerian disk,
corona, and central jet---starting from nothing but a simple constant
angular momentum torus.  Many of these features can be understood analytically,
at least in a crude sense.

\item Energetics (including radiation and dissipative physics) remains
in its infancy--we know less than we think.  Numerical disk models have
yet to be placed on the observational plane.

\item We have not tested under what conditions the fundamental formula
of phenomenological disk theory $$ Q_- = - T_{R\phi}\, {d\Omega\over
d\ln R}$$ is valid.  Even the eclipse mapping observations that show an
$R^{-3/4}$ power law dependence in the temperature profile of CV disks
do not require this assumption to be valid, though they are certainly
consistent with it.  Investigating this question numerically should be
a central goal for numericists.

\item A closely related point:
We are completely ignorant of the role of thermal energy
transport, $\delta v \, \delta T$.  It need not be small relative to
rotational transport, and can be extracted from simulations.

\item The consequences for the MRI of departures from ideal MHD in
the form of Hall EMFs and ohmic resistance are critical to understand.
Only one team, Sano \& Stone, has studied this numerically, and then
only on fairly coarse grids.  Resolution may be key here.  The question
that needs to be thoroughly explored is under what circumstances does
the MRI turn of?

\item The temperature dependence of the resistivity has not been touched
numerically.  Clearly, there is a real potential to explore eruptive
behavior here.

\item For all their foibles and limitations, numerical MRI simulations have
taken their place at the helm of theoretical accretion disk studies,
and they are likely to remain there for the foreseeable future.

\end{itemize}

\section*{Acknowledgements}
I thank J.-P. Lasota for helpful comments on the manuscript.
Support from NASA grants NAG5-13288 and NNG04GK77G is gratefully
acknowledged.

{}

\end{document}